\documentclass[12pt]{article}
\usepackage{epsf}
\def\beq{\begin{equation}}
\def\eeq{\end{equation}}

\begin{document}
\begin{titlepage}
\begin{center}
{\Large \bf William I. Fine Theoretical Physics Institute \\
University of Minnesota \\}
\end{center}
\vspace{0.2in}
\begin{flushright}
FTPI-MINN-05/46-T \\
UMN-TH-2418-05 \\
ITEP-TH-63/05\\
November 2005 \\
\end{flushright}
\vspace{0.3in}
\begin{center}
{\Large \bf  Particle decay in false vacuum
\\}
\vspace{0.2in}
{\bf A.~Gorsky} and {\bf M.B.~Voloshin  \\ }
William I. Fine Theoretical Physics Institute, University of
Minnesota,\\ Minneapolis, MN 55455 \\
and \\
Institute of Theoretical and Experimental Physics, Moscow, 117259
\\[0.2in]
\end{center}

\begin{abstract}
We revisit the problem of decay of a metastable vacuum induced by the presence
of a particle. For the bosons of the `master field' the problem is solved in any
number of dimensions in terms of the spontaneous decay rate of the false vacuum,
while for a fermion we find a closed expression for the decay rate in (1+1)
dimensions. It is shown that in the (1+1) dimensional case an infrared problem
of one-loop correction to the decay rate of a boson is resolved due to a
cancellation between soft modes of the field. We also find the boson decay rate
in the `sine-Gordon staircase' model in the limits of strong and weak coupling.
\end{abstract}

\end{titlepage}

\section{Introduction}
The decay of a metastable vacuum state is a quite universal problem in quantum
field theory. The decay proceeds through nucleation and subsequent classical
expansion of the bubbles of the true vacuum. The classical bubbles can exist
only starting from a certain critical radius at which the energy loss due to the
surface terms is compensated by the gain in the volume energy. The formation of
the critical bubbles is thus a quantum tunneling process\cite{vko} which
tunneling can be described by an Euclidean-space configuration of the field,
called a `bounce'\cite{cc1}. The space-time nucleation rate of the critical
bubbles, $w_0$, i.e. the probability of such nucleation per unit time and per
unit volume, is proportional to the exponent of the classical action on the
bounce configuration: $w_0 \propto \exp(-S_{cl})$, while the pre-exponential
factor requires a calculation of the functional determinant at the bounce. The
exponential factor is readily found\cite{vko,cc1} in the so called thin wall
limit, namely when the radius of the critical bubble is much bigger than the
effective thickness of its wall. This limit is always realized at a small
difference $\epsilon$ between the vacuum energy density of the false and the
true vacua, with the other parameter determining $S_{cl}$ being the surface
tension $\mu$ of the bubble wall, i.e. of the boundary between the metastable
and the stable phases. The pre-exponential factor in the bubble nucleation rate
is known in a closed form only in (1+1) dimensional models\cite{ks,mv}, where in
the thin wall limit it is determined only by the parameter $\epsilon$, with only
partial results found in (2+1) dimensions\cite{garriga,mr,mv2}, and virtually no
result known in the (3+1) dimensional case.

Similarly to the behavior in the decay of a metastable phase in a thermal
setting, the presence of matter in the false vacuum generally provides `centers
of nucleation' for the bubbles of the true vacuum. Thus one can consider the
false vacuum decay induced by the presence of a particle\cite{adl,sv,mv3}, by
particle collisions\cite{sv,mv3}, by matter with finite density\cite{gk}, as
well as by the matter being in a thermal equilibrium where the problem goes back
to the more conventional thermodynamic setting\cite{langer}. In this paper we
revisit the calculation of the bubble nucleation rate associated with the
presence of a particle in the false vacuum. The particle-induced nucleation can
also be viewed as the decay of the particle (albeit in the process the initial
`vacuum' state also gets destroyed), whose rate $\Gamma$ generically can be
written in the form $\Gamma= K \, w_0$, where the constant $K$, which can be
naturally called the `catalysis factor', is the main subject of our
consideration. The catalysis is most efficient for the particles which have zero
modes localized on the boundary between the false and the true vacua. The reason
for this behavior is that in this case the energy corresponding to the mass of
the particle $m$ in the initial state is fully transferred to the bubble degrees
of freedom, since in the final state the particle ends up as a zero mode
localized on the bubble wall. This effectively corresponds to the upward shift
by $m$  of the energy at which the tunneling takes place\cite{sv}, and results
in $K$ being proportional to the exponential factor $\exp( 2 \, m \, \tau)$,
where $\tau$ is the (Euclidean) time on the tunneling trajectory. The
exponential behavior due to the shift of the energy for the tunneling trajectory
can be found explicitly both in (1+1) dimensions\cite{sv} and in
higher-dimensional models\cite{mv3}. However the pre-exponential factor has been
calculated only for the bosons of the master field in a (3+1) dimensional
case\cite{adl}, for which bosons the existence of the zero mode is always true.
Here we calculate the pre-exponential behavior of the catalysis factor for the
same bosons in lower dimensions, and also find a closed formula for this factor
in (1+1) dimensions for a fermion, whose field has a zero mode on the
inter-vacua boundary. The existence of such fermionic mode is a generic
phenomenon and is guaranteed in the case where the mass term for the fermions
changes sign across the bubble wall\cite{jr}.

Our consideration, similarly to Ref.\cite{adl}, is generally limited to models
with weak coupling, which implies that the masses $m$ of the both types of
considered particles are small in comparison with the scale of the surface
tension $\mu$. In this case the deformation of the tunneling trajectory due to
the energy shift by $m$\cite{sv,mv3} can be neglected, so that, in particular,
the tunneling time $\tau$ coincides with the radius $R \propto \mu/\epsilon$ of
the critical bubble, $\tau=R$, as it does in the spontaneous vacuum
decay\cite{cc1}. Furthermore, we also assume the applicability of the thin wall
limit, which implies the condition $m \, R \gg 1$, and which is always valid in
the limit of small $\epsilon$.

The catalysis factor $K$, as defined, has the dimension of the spatial volume.
Thus it would be natural to compare the pre-exponential factor in $K$ with the
spatial volume of the critical bubble of the radius $R$. Under our assumptions
we find that for a fermion in (1+1) dimensions this factor is indeed of order
$R$, while the catalysis factor for the bosons is enhanced in comparison with
the volume of the bubble by inverse powers of the (small) coupling constant.

It should be noted that technically the bosonic catalysis factor in lower
dimensions can be found by a straightforward application of the treatment of
Ref.\cite{adl}. Such application is fully justified in a (2+1) dimensional case.
However in (1+1) dimensions there is a potential complication in estimating the
effect of the quantum fluctuations arising from an infrared behavior of the
modes of the bosonic field over the bounce background. We demonstrate for this
case that the large infrared terms in fact cancel due to the specific properties
of the soft modes.

The material in the rest of the paper is organized as follows. In Sec.2 we
briefly review the calculation of the spontaneous decay rate of false vacuum and
present a calculation of the decay rate induced by a boson of the scalar field,
which defines the vacuum states. In Sec.3 the problem of the infrared behavior
of the one-loop correction to the calculated decay rate in (1+1) dimensions is
considered and it is shown that this problem is resolved due to a cancellation
of the contributions to this correction between the negative mode and the sum
over the positive soft modes of the field of the bounce. In Sec.4 the catalysis
factor is calculated for a fermion in a (1+1) model. We then discuss the decay
of metastable states in the sine-Gordon model with added linear term, the so
called `sine-Gordon staircase'. Using the equivalence\cite{stone} of this
bosonic model and the massive fermionic Thirring model in an external electric
field, we find the induced decay rate for both the weak coupling limit (Sec.5)
and for the strong coupling limit (Sec.6), the latter corresponding to a weak
coupling in the Thirring model. Finally, in Sec.7 we discuss possible
implications of our calculation for other models.

\section{Spontaneous and induced decay of false vacuum}
In what follows we assume a situation where the energy density of a scalar field
$\phi$, the `master field', has a local minimum at $\phi=\phi_+$, which is
higher than in a neighboring minimum at $\phi=\phi_-$. The vacuum state defined
by the former minimum is referred to as the false vacuum, while the latter is
the true vacuum. A typical example of such situation is provided by the well
known model of a scalar field with the potential
\beq
V(\phi)={\lambda^2 \over 8} \, \left ( \phi^2 - v^2 \right )^2 + a \, \phi~,
\label{pot}
\eeq
where $\lambda$, $v$, and $a$ are constants. At $a=0$ the potential has two
degenerate minima at $\phi_\pm = \pm v$, while at small positive $a$ the
degeneracy is lifted in such a way that the minimum at $\phi_+$ has energy
density bigger than that of $\phi_-$ by the amount $\epsilon \approx 2\, a\, v$.
The vacuum state at $\phi_+$, being stable at $a \le 0$ becomes metastable at
positive $a$ and decays by nucleation and subsequent expansion of the bubbles
filled with the phase $\phi_-$. At small $a$ the surface density of the bubble
wall can be approximated\cite{vko,cc1} by the surface density of the soliton
with the field profile
\beq
\phi(x) = v \, \tanh {m \, x \over 2}
\label{wall}
\eeq
interpolating between the two degenerate vacua in the limit $a \to 0$:
\beq
\mu=\int_{-v}^{+v} \, \sqrt{2 V(\phi)} \, d\phi = {2 \over 3}\, \lambda \, v^3~.
\label{mu}
\eeq
The mass of the scalar particles of the field $\phi$ propagating in either of
the vacua is given (also in the limit $a \to 0$) as $m=\lambda \, v$. In a model
with the total space-time dimensions equal to $d$ the ratio $m^{d-1}/\mu$
coincides with the dimensionless coupling constant for the perturbation theory
in this model. We assume throughout this paper that this ratio is a small
parameter, which thus corresponds to weak coupling.

In the Euclidean-space formulation of the problem of the false vacuum
decay\cite{cc1,cc2} the calculation of the spontaneous decay rate amounts to a
semiclassical evaluation of the imaginary part of the energy of the false vacuum
from the path integral
\beq
Z= {\cal N} \int \, e^{-S[\phi,\ldots]} \, {\cal D}\phi \ldots
\label{funint}
\eeq
where the dots stand for other possible fields present in a specific model,
${\cal N}$ is the normalization factor, and the integration is performed with
the condition that the field $\phi$ approaches its false vacuum value $\phi_+$
at the boundaries of the space-time normalization box. The decay rate is then
given by $w_0 = 2 \, {\rm Im} (\ln Z)/VT$, where $VT$ is the space-time volume
of the normalization box.

The action functional $S$ has a semiclassical saddle point at the configuration
described by the bounce\cite{cc1}. In the thin wall limit the bounce is an
$O(d)$ symmetric bubble with the field $\phi_-$ inside and $\phi_+$ outside, and
the bubble wall, separating the two phases has the surface tension $\mu$. The
action for the bounce in this approximation is given by
\beq
S=\mu \, A_B - \epsilon \, V_B~,
\label{act}
\eeq
where $V_B$ is the $d$ dimensional volume of the bounce and $A_B$ is its $(d-1)$
dimensional surface area. The action (\ref{act}) reaches its extremum on a
spherical bounce with the radius $R=(d-1)\, \mu/\epsilon$, which is also the
radius of the critical bubbles capable of classical expansion in the Minkowski
space-time.

The spectrum of small deformations of the bounce around the extremum contains
exactly one negative mode, corresponding to an overall variation of the radius.
This mode in fact gives rise to the imaginary part\cite{cc1} of the path
integral in eq.(\ref{funint}). Furthermore this spectrum also contains $d$
translational zero modes, the integration over which introduces the factor of
the space-time volume $VT$ in the contribution of the bounce to the energy of
the vacuum state.


The decay rate of a particle of the field $\phi$ in the false vacuum can be
calculated\cite{adl,sv} by considering the imaginary part of the contribution of
a bounce to the Euclidean-space propagator of the excitations
$\sigma(x)=\phi(x)-\phi_+$ of the field $\phi$:
\beq
D(x,y)={1 \over Z} \, \int \, \sigma(x) \sigma(y)\, e^{-S[\phi,\ldots]} \, {\cal
D}\phi \ldots
\label{prop}
\eeq
in the limit of large separation $L=|x-y|$. Indeed, the contribution of the
bounce to the correlator (\ref{prop}), as shown in Fig.1a, has the generic form
\beq
\delta D(x,y) = {i \over 2}\, w_0 \int d^dz \, F(x-z,y-z)\, D_0(x-z)\, D_0
(y-z)~,
\label{dprop}
\eeq
where $D_0(x)$ is the free-particle propagator in the vacuum $\phi_+$,
satisfying the equation
\beq
\left( - \partial^2 +m^2 \right ) D(x) = \delta^{(d)}(x)
\label{prop0}
\eeq
and the factor $(i/ 2)\, w_0  d^dz$ is the proper measure of integration over
the coordinate $z$ of the center of the bounce, as follows from the
consideration of the bounce contribution to (the imaginary part of) the vacuum
energy.
\begin{figure}[ht]
\begin{center}
 \leavevmode
    \epsfxsize=12cm
    \epsfbox{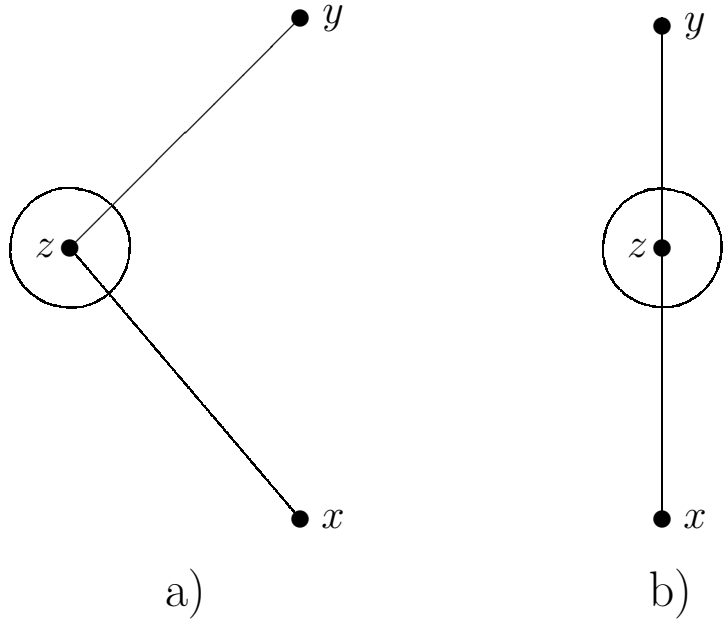}
  
    \caption{The configurations for the bounce contribution to the propagator
(eq.(\ref{dprop})). A generic configuration ($a$) and the alignment of the
bounce position ($b$), dominating the integral in eq.(\ref{dprop}) at large
$|x-y|$. }
\end{center}
\end{figure}

Let us consider the contribution to the integral (\ref{dprop}) arising from the
configurations, where the bounce is far (in units of its radius) from either of
the points $x$ and $y$, i.e. where $|x-z| \gg R$ and $|y-z| \gg R$. The
propagators $D_0(x-z)$ and $D_0 (y-z)$ in the integral in eq.(\ref{dprop})
describe the exponential attenuation of the correlation  ($D(x) \sim \exp(-m
|x|)$ at large separations, while the form factor $F(x-z,y-z)$ does not have
this exponential behavior. For this reason at $|x-y|=L \gg R$ the integrand in
eq.(\ref{dprop}) is maximized for $z$ lying on the straight line running between
$x$ and $y$: $z_\nu=s \, (x_\nu-y_\nu)/L$, and the integration can be split into
the longitudinal, over the parameter $s$ along this line, and the transversal,
over $z_\perp$. The integration over $z_\perp$ can be done by the saddle point
method, so that the form factor $F(x-z,y-z)$ can be replaced by its value at
$z_\perp = 0$, and the essential configuration to be considered is the one shown
in Fig.1b. As will be discussed few lines below, when the bounce is far from the
endpoints of integration over $s$, i.e. $s \gg R$ and $L-s \gg R$, the value of
the form factor in fact does not depend on $s$ and is a constant $F_0$. Since
the contribution of the excluded regions around the endpoints is only of
relative order $R/L$, the integral in eq.(\ref{dprop}) can be replaced at large
$L$ by
\beq
\delta D(x,y) = {i \over 2}\, w_0 \, F_0 \, \int d^dz \, D_0(x-z)\, D_0 (y-z)~,
\label{dprop1}
\eeq
where $F_0$ should be calculated from the configuration shown in Fig.1b.

The expression (\ref{dprop1}) for the modification $\delta D$ of the propagator
by the bounce can be compared with the first-order correction to the propagator
due to a small shift of mass by $\delta m^2$, $m^2 \to m^2 + \delta m^2$, in
eq.(\ref{prop0}). In the standard way one finds
\beq
\delta_m D(x,y) = -\delta m^2 \, \int d^dz \, D_0(x-z)\, D_0 (y-z)~.
\label{dpropm}
\eeq
Thus the contribution (\ref{dprop1}) of the bounce to the propagator of the
boson in the false vacuum is equivalent to an imaginary shift of the boson mass:
$\delta m^2= - (i / 2)\, w_0 \, F_0$, which corresponds to the particle decay
rate given by $\Gamma= F_0 w_0/(2m)$, so that the catalysis factor $K$ is found
as
\beq
K = {F_0 \over 2 m} ~.
\label{gamb}
\eeq

The factor $F_0$ can be readily found\cite{adl} for the discussed here case of
the bosons of the classical field of the bounce. Indeed, consider the classical
field just outside the bounce, i.e. at the distance $r > R$ from the center,
such that $r-R \gg m^{-1}$, but still $r-R \ll R$. The former condition ensures
that the field is described by its asymptotic approach to the vacuum value
$\phi_+$, while the latter implies that in this region the curvature of the
bounce wall can be neglected in a calculation of this asymptotic behavior. Thus
one can consider instead the asymptotic behavior of the field in the limit
$\epsilon \to 0$, i.e. of the field of the stable soliton separating two
degenerate vacua. This asymptotic behavior has the form $\phi(x)-\phi_+ = - 2 v
\exp[-m \, (r-R)]$, where in the model described by the potential (\ref{pot})
$v$ coincides with the corresponding parameter in the potential, while in a
generic model $v \sim (\phi_+ - \phi_-)/2$. On the other hand in the
$O(d)$-symmetric problem the asymptotic approach of the scalar field to its
vacuum value is described by the solution of the linearized
spherically-symmetric equation, equivalent to the homogeneous part of
eq.(\ref{prop0}),  and reads as,
\beq
\phi(r)-\phi_+ = C \, D_0(r)~
\label{defc}
\eeq
where the free boson propagator in $d$ dimensions has the well known expression
in terms of the modified Bessel function $K_\nu(mr)$:
\beq
D_0(r)= {m^{d/2-1} \over (2 \pi)^{d/2} \, r^{d/2-1}} K_{d/2-1}(mr)~.
\label{d0}
\eeq
The constant $C$ in the asymptotic expression (\ref{defc}) is found by comparing
the two expression for $\phi(x)-\phi_+$ in the discussed region just outside the
bounce and using the standard asymptotic formula for the function $K_\nu(mr)$.
In this way one finds
\beq
C= - 4 \, (2 \pi)^{d/2-1} \, m^{(3-d)/2} \, R^{(d-1)/2} \, v \, e^{m \, R}~.
\label{cexp}
\eeq
Using then the expression (\ref{defc}) for the field with thus determined
constant $C$, one finds the product of the classical fields $\sigma(x)
\sigma(y)$ in the integral in eq.(\ref{prop}) in the configuration shown in
Fig.1b, corresponding to the constant $F_0$ in eq.(\ref{dprop1}) given by
\beq
F_0 = C^2 = 16 \, (2 \pi)^{d-2} m^{3-d} \, R^{(d-1)} \, v^2 \, e^{2 \, m \, R}~,
\label{fres}
\eeq
which indeed does not depend on the position of the bubble along the straight
line connecting the points $x$ and $y$ as long as both these points are
sufficiently outside the bounce. The catalysis factor thus can be found from the
relations (\ref{gamb}) and (\ref{fres}) in the form
\beq
K=2^{d+1} \, \pi^{(d-3)/2} \, \Gamma \left({d+1 \over 2} \right ) \, m^{2-d} \,
v^2 \, V_{d-1} \, e^{2 \, m \, R}~,
\label{catb}
\eeq
where $V_{d-1}= \pi^{(d-1)/2} \, R^{d-1}/ \Gamma[(d+1)/2]$ is the spatial ($d-1$
dimensional) volume of the critical bubble. As discussed in the Introduction, it
is natural to compare the catalysis factor with this volume. The result in
eq.(\ref{catb}) shows that besides the classical exponential factor the
catalysis is additionally enhanced by the factor $ m^{2-d} \, v^2$ in the
pre-exponent, which is the inverse of the small dimensionless coupling in the
theory.

\section{Boson-induced decay in (1+1) dimensions}

The formula for the catalysis factor in eq.(\ref{catb}) reduces in (3+1)
dimensions to the result of Ref.\cite{adl}, and in other dimensions it presents
a rather straightforward generalization. There is however one point, of a
special importance to a (1+1) dimensional case, related to the effect of the
quantum fluctuations on the essentially classical result in eq.(\ref{catb}).
Generally, the effect of the quantum fluctuations (the loop correction) is
expected to be suppressed by a power of the coupling constant as compared to the
classical contribution. In the discussed problem this expectation is true at $d
>2$, however in a (1+1) dimensional problem this expectation is potentially
jeopardized by an infrared behavior. Indeed the eigenvalues of the second
variation of the action for the fluctuations of the shape of the bounce,
described by the effective action (\ref{act}), are proportional to $R^{-2}$. The
modes with these eigenvalues are localized on the bounce boundary and describe
the soft part of the spectrum of the modes of the field around the stationary
bounce configuration, as opposed to the modes, whose eigenvalues start at
$O(m)$, and those `hard' modes describe the excitations propagating in the bulk
as well as possible deformations of the profile of the field across the bounce
wall. Let us estimate the contribution of an individual soft mode with the
eigenvalue $c_n/R^2$ to the correlator (\ref{prop}), with $c_n$ being a number.
All such modes originate from local shifts of the wall of the bounce, so that
the field profile of an individual mode is proportional to the radial derivative
of the field of the bounce, $\phi'(r)$. The field $\sigma_n$ of a normalized to
one mode in (1+1) dimensions is then parametrically estimated at the distance $r
> R$, such that $r-R \gg m^{-1}$, but still $r-R \ll R$, as
\beq
\sigma_n(r) \sim {m \,  v \over \sqrt{\mu R}} \, e^{-m (r-R)} \sim \sqrt{m \over
R}\, e^{-m (r-R)} ~,
\label{sest}
\eeq
where it is taken into account that $\int(\phi')^2 \, dr \approx \mu \sim m \,
v^2$, and any numerical factors are dropped for a parametrical estimate. The
contribution of such mode to the correlator (\ref{prop}) is then proportional to
\beq
{R^2 \over c_n} \, \sigma_n(r_1) \sigma_n(r_2) \sim c_n^{-1} \,  (m \, R) \,
e^{2 m R} \, e^{-m (r_1+r_2)}~.
\label{s2est}
\eeq
In this estimate the large factor $(m \, R)$ stands in place of the factor $v^2$
in the similar product for the classical field. Thus the magnitude of an
individual mode contribution to the correlator relative to the classical part is
described by the parameter
$(m \, R)\over v^2$, which in-spite of the expected suppression by the
dimensionless coupling $v^{-2}$ is infrared unstable at large $R$.

We will show however that the sum over the soft modes gives zero for this
infrared contribution due to a cancellation between one negative and all
positive modes. In order to demonstrate this we consider the parametrization of
the shape of the bounce in the polar coordinates $(r, \theta)$ on a plane, so
that the effective action (\ref{act}) for the soft modes takes the form
\beq
S=\int_0^{2\pi} \left ( \mu \, \sqrt{r^2 + {\dot r}^2} - {1 \over 2} \, \epsilon
\, r^2 \right ) \, d\theta = {\pi \, \mu^2 \over \epsilon} + \int_0^{2\pi} \,
{\epsilon \over 2} \, ({\dot \rho}^2 - \rho^2) \, d \theta + O(\rho^4),
\label{act2}
\eeq
where the latter expression shows the two first terms of expansion in the small
deviation $\rho=r-R$ of the radial variable $r$ from its stationary value
$R=\mu/\epsilon$, and the dot stands for the derivative over $\theta$. The
quadratic part in this expression has one negative eigenmode $\rho = 1/\sqrt{2
\pi}$ and the spectrum of zero and positive double degenerate eigenmodes:
\beq
\rho_n^{(1)} = {1 \over \sqrt{\pi}}\, \cos n \theta~,~~~ {\rm and} ~~~
\rho_n^{(2)} = {1 \over \sqrt{\pi}} \, \sin n \theta~; ~~~~~(n=1,2,\ldots)~.
\label{modes}
\eeq
The spectrum of the eigenvalues is proportional\footnote{The proportionality
coefficient is not important for this discussion. It can be noted however that
in terms of the normalized modes for the field $\phi$ the eigenvalues are
$(n^2-1)/R^2$.} to $(n^2-1)$ with the negative mode corresponding to $n=0$.

Let us consider now the configuration shown in Fig.1b with the bounce located on
the line connecting the points $x$ and $y$. Let the angle $\theta$ be defined as
measured counterclockwise from the downward vertical connecting the center of
the bounce with the point $x$, so that $\theta = \pi$ corresponds to the upward
vertical connecting the same center with the point $y$. Clearly, the
contribution of the fluctuations of $\rho$ to the propagator (\ref{prop}) is
proportional to
\beq
\langle [\rho(0)+\rho(\pi)]^2 \rangle \propto \sum_n {[\rho_n(0)+\rho_n(\pi)]^2
\over n^2 -1}~.
\label{sum0}
\eeq
Note however that the sum $\rho(0)+\rho(\pi)$ is not vanishing only for the
negative mode and for the positive modes of the first type, $\rho_n^{(1)}$, with
even $n$, i.e. $n=2 k$. Thus the sum in eq.(\ref{sum0}) is proportional to
the numeric sum
\beq
-{1 \over 2} + \sum_{k=1}^\infty {1 \over 4 \, k^2 -1} =0~,
\label{sum2}
\eeq
where the first term is due to the negative mode and the sum runs over the
positive modes.  The arithmetic identity (\ref{sum2}) explicitly demonstrates
that the infrared contribution in a (1+1) dimensional model cancels between the
negative mode and the sum over the positive ones.

Let us also remark on the decay of a moving particle in the false vacuum.  If
particle moves with a constant
velocity then the probability depends on the velocity through the standard
Lorentz factor. However if it
moves with the constant acceleration situation is more subtle since in the
particle frame
vacuum behaves as the thermal bath due to Unruh effect. The effective
temperature is
defined through the acceleration as
\beq
T_{eff}={a \over 2\pi}
\eeq  
hence probability of the particle decay is modulated by the thermal effects. The
most
essential effect corresponds to the possible deformation of the classical
bounce. Since
temperature corresponds to the periodicity in the Euclidean time then the
deformation
of the bounce happens when period corresponding to the temperature becomes
comparable to  $2R$. That is   deformation
of the bubble emerges if the acceleration is larger then
\beq
a_{crit}={\pi \epsilon \over \mu}.
\eeq
and our approximation fails.

\section{Fermion-induced decay in (1+1) dimensions}
The very existence of fermions in a model is known to modify (without any
fermions being present in the initial state) the pre-exponential factor in the
rate of the false vacuum decay in the situation where the complex fermion field
$\psi$ has a zero mode on the boundary between the vacua (in the limit $\epsilon
\to 0$). Such situation takes place when the mass term for the fermion changes
sign between the two vacua\cite{jr}. In particular the rate $w_0$ for the
spontaneous decay of the false vacuum in (1+1) dimensions receives a factor of 2
in comparison with purely bosonic theory\cite{ks2,mv4}. This doubling
corresponds to the existence of two final states in the false vacuum decay in
(1+1) dimensions viewed as a spontaneous creation of a kink-antikink pair: one
state where both the kink and the antikink are created with the fermion zero
mode empty, and the other state is where is a zero-energy fermion on the kink
and a zero-energy antifermion on the antikink.

In what follows we assume that the interaction of the fermion field with the
scalar field of the bounce is such that there exists a zero fermion mode on the
kink separating the two vacua.
In order to find the effect of a fermion on the probability of nucleation of a
critical bubble we consider the bounce contribution to the fermion propagator
$G(x,y)=\langle \psi(x) {\overline \psi} (y)\rangle $ in the configuration shown
in Fig.1b. Clearly, the exponentially enhanced factor $\exp(2 \, m_f \, R)$,
where $m_f$ is the mass of the fermion, arises from the contribution of the zero
mode, whose propagation from the lower point of the bounce to the upper one does
not contain any exponential attenuation. Assuming for definiteness that the
fermion mass is positive (and equals $m_f$) in the false vacuum, and choosing
the $\gamma$ matrices as $\gamma_1=\sigma_1$ and $\gamma_2=\sigma_2$, one finds
the solution for the Dirac equation for the field of the zero mode of $\psi$ in
the background scalar field $\phi(r)$ of the bounce,
\beq
\left [ \sigma_i \partial_i +m(\phi)  \right ] \psi_0 = 0~,
\label{dirac}
\eeq
in the form
\beq
\psi_0(r, \theta)=C_f \, \sqrt{R \over r} \, \exp \left \{ -\int_R^r \,
m[\phi(r')] \, dr' \right \} \chi(\ell) \, \left (\begin{array}{c} e^{-i \,
\theta/2 } \\  e^{i \, \theta/2 } \end{array} \right)~,
\label{psi0}
\eeq
where $\chi(\ell)$ is a one-dimensional fermion field living on the bounce
boundary and (nominally) depending on the length parameter $\ell = R \theta$
along the boundary. Notice that the classical equation for $\chi$ reads $\dot
\chi =0$. Finally, the constant $C_f$ in eq.(\ref{psi0}) is the normalization
factor relating the normalization of $\psi$ and $\chi$ and satisfying the
condition
\beq
2 \, C_f^2 \, \int \exp \left [ -2 \int_R^r \, m(\phi) \, dr' \right]\, dr = 1~.
\label{cfact}
\eeq
In what follows we rather use a  related factor ${\tilde C}_f$ defined as the
coefficient in the expression
\beq
C_f \, \exp \left \{ -\int_R^r \, m[\phi(r')] \, dr' \right \} \approx {\tilde
C}_f \, \exp \left [ m_f (R-r) \right ]~,
\label{tcfact}
\eeq
which is valid sufficiently far outside the bounce where $m_{min} \, (r-R) \gg
1$ with $m_{min}$ being the minimal mass scale in the model. Generally the
factor ${\tilde C}_f$ can be estimated as
\beq
{\tilde C}_f^2 = {m_f \over 2} \, f\left ( {m_f \over m} \right )~,
\label{mrat}
\eeq
where $f$ is a dimensionless function of the ratio of $m_f$ to masses of other
particles in the false vacuum. In the limit where $m_f$ is much smaller than
other masses one has $f(0)=1$. In the model, where the scalar `master field' is
described by the potential (\ref{pot}) and the mass of the fermion is
proportional to $\phi$, the function $f$ can be found explicitly:
\beq
f(u)={2^{2 u} \over \sqrt{\pi}} \, {\Gamma(u+1/2) \over \Gamma(u+1)}~.
\label{funcf}
\eeq

The contribution of the fermion zero mode on the bounce the configuration shown
in Fig.1b can be written, using the asymptotic behavior of the zero mode
(\ref{psi0}), in terms of the one-dimensional propagator of the field $\chi$ on
the boundary, $g(\ell_1, \ell_2)=\langle \chi(\ell_1) \chi^\dagger(\ell_2)
\rangle$, as
\beq
\delta G(x,y)=-{i  \over 2} \, {{w_0} \over 2} \, {d^2z} \,  {\tilde C}_f^2 \,
e^{2 \, m_f \, R} \, R \ {e^{-|x-y|} \over \sqrt{|x-z| \, |y-z|}} \,
(1+\sigma_1) \, g(0,\pi R) ~.
\label{delg}
\eeq
Notice that for a complex fermion there is only one path for propagation along
of $\chi$ along the boundary from the bottom of the bounce to its top (assumed
here for definiteness to be counterclockwise in terms of Fig.1b), corresponding
to the final state, where the fermion is a bound state localized on the kink.
The other path (clockwise) would be relevant for the vacuum decay induced by an
anti-fermion, which in the final state is localized on the antikink. The
expression in eq.(\ref{delg}) contains an extra factor 1/2, due to the fact
that, as mentioned before, the spontaneous nucleation rate $w_0$ in the theory
with fermions contains extra factor of 2, due to the existence of two final
states in the decay, which is to be compensated in the proper measure of
integration over the coordinate of the center of the bounce $d^2z$. The
propagator $g$ has a very simple explicit form in terms of the sign function:
$g(\ell_1,\ell_2)= (1/2) \, {\rm sign(\ell_1- \ell_2)}$, so that $g(0,\pi
R)=-1/2$.

The expression in eq.(\ref{delg}) can now be compared with the corresponding
change of the free propagator $G_0$ under a shift $\delta m_f$ of the fermion
mass:
\beq
\delta_m G(x,y)=-\delta m_f \, d^2 z \,  G_0(x-z) \, G_0(z-y) \to -\delta m_f \,
d^2 z \, {m \over 4 \pi} \, (1+ \sigma_1) \, {e^{-|x-y|} \over \sqrt{|x-z| \,
|y-z|}} ~,
\label{delmg}
\eeq
where the asymptotic expression takes into account the explicit form of the free
propagator:
\beq
G_0(x,y)={1 \over 2 \pi} \,(-\sigma_i \partial_i +m ) \,  K_0(m_f \, |x-y|)~.
\label{g0}
\eeq
Using this comparison and eq.(\ref{mrat}) one finds the imaginary part of the
fermion mass shift corresponding to the decay rate of the fermion
\beq
\Gamma_f= {\pi \over 2} \, f\left ( {m_f \over m} \right ) \, R \, w_0 \, \exp
(2 \, m_f \, R)  = {\mu \over 2} \, f\left ( {m_f \over m} \right ) \, \exp
\left ( -{\pi \, \mu^2 \over \epsilon} + 2 \, m_f \, R \right )~.
\label{ferg}
\eeq
Here in the latter transition are used the explicit expressions: $w_0 =
(\epsilon/\pi) \, \exp(-\pi \, \mu^2/\epsilon)$ and $R = \mu/\epsilon$. One can
readily see that in the fermion case, as expected, the pre-exponent in the
catalysis factor, $K_f= (\pi/ 2) \, f(m_f/m) \, R \, \exp (2 \, m_f \, R)$, is
indeed of the order of the spatial size of the critical bubble.

\section{Meson decay in weakly coupled sine-Gordon model}

In order illustrate the universality of the derived results let us discuss
one more example
where the particle decay in the false vacuum happens in two dimensions.
We shall derive the probability of the decay of the electrically neutral
meson bound state in the Thirring model with the pre-exponential
accuracy.

Consider the sine-Gordon theory with the Lagrangian
\beq
L_{SG}={1 \over 2} \, (\partial \phi)^2 +\frac{\alpha}{\beta^2} \cos (\beta
\phi)
\eeq
and add term $(\epsilon \, \beta/2\pi)\,  \phi$ which yields the situation
with the metastable states. This theory upon the two-dimensional
fermionization is equivalent to the massive Thirring model with the
Lagrangian
\beq
L_{Th}= i\bar{\psi}\partial_{\nu}\gamma^{\nu}\psi -{1\over 2}\,
g\,j^{\nu}j_{\nu} + \mu \, \bar{\psi}\psi +
A_0 \,j_0
\eeq
where ${{\beta}^2\over 4\pi}=(1+{g\over  \pi})^{-1} $, and $j_\nu=
\bar{\psi}\gamma_{\nu}\psi$.
One can identify $\mu$ with the soliton mass in the sine-Gordon model,
and $\partial_{x} A_0=\epsilon$.  In what follows we shall
assume that $\beta^2< 4\pi$ which is the condition
for the bound state of fermions to exist in the Thirring model.  The solitons
in the sine-Gordon model get mapped into the fermions in the Thirring model
while the field $\phi$ gets mapped into the fermion-antifermion meson bound
state.

The linear perturbation term in the sine-Gordon model corresponds to the
constant electric field
in the Thirring model  realization, so that the problem of false vacuum decay
can be discussed in
both formulations. In the Thirring model it corresponds to the Schwinger
pair production. The probability of the spontaneous vacuum decay in the
sine-Gordon
model and equivalent Schwinger process in Thirring model has been found in
\cite{stone}. The one-loop result coincides with the general formula
$w_{Th} = (\epsilon/2\pi) \, \exp(-\pi \, \mu^2/\epsilon)$, while in the special
case of $\beta^2=4\pi$, corresponding to $g=0$, the exact result is found as
\beq
w_{Th} =-{\epsilon\over  2 \pi} \, \ln(1 -e^{- \pi \mu^2 / \epsilon})
\eeq
Note that in this case the bose-fermi equivalence allows to perform
the summation over the multiple bounces in the sin-Gordon theory.

Now we can discuss the decay of the false vacuum in the
presence of a particle corresponding to the field $\phi$ in the sine-Gordon
model. In the weak coupling regime for the bosons, i.e. at small $\beta$, the
soliton is much heavier
then the boson particle, which corresponds to the  situation
where the external particle does not deform the classical
bounce configuration.
Hence the decay rate can be immediately read off eq.(\ref{catb}).
In this case the process corresponds  in the Thirring model to the
nonperturbative decay
of the light electrically neutral meson in the electric field and
the catalysis factor  of this process  is
\beq
K_{Th} = {32 \over \beta^2} \, {\mu \over \epsilon} \, e^{{2 \, m_{b} \mu /
\epsilon}}
 \eeq
where $m_b$ is the meson mass.
Note that this process is the two-dimensional counterpart
of the induced Schwinger processes discussed in four dimensions
in \cite{gss,monin} in the exponential approximation.

\section{Meson decay in strongly coupled sine-Gordon model}
The boson-fermion correspondence in this model actually allows to find the meson
decay rate in the limit, opposite to what has been considered so far in this
paper, namely for strongly coupled bosons, when the boson mass is close to the
kink-antikink threshold. This limit corresponds to a small positive $g$, and the
boson mass (at $\epsilon \to 0$) is $m_b=2\,\mu - \mu \, g^2$. The
near-threshold dynamics of the soliton-antisoliton pair can be considered
nonrelativistically as a motion of a pair with the reduced mass $\mu/2$ in the
local potential $U(x)=-2 g \, \delta(x)$, which correctly reproduces the energy
of the bound state (the boson). For a nonzero $\epsilon$ the nonrelativistic
Hamiltonian for this system takes the form:
\beq
H={p^2 \over \mu} - \epsilon \, x -2 g \, \delta(x)~.
\label{hamnr}
\eeq
The problem of the boson decay in the false vacuum is reduced in terms of this
equivalent nonrelativistic system to that of ionization of the bound state in
the external electric field $\epsilon$.

In order to solve the ionization problem we start with considering the Euclidean
time propagator (``the heat kernel") defined as
${\cal K}(x, y; \tau)= \langle x | \exp(-H \, \tau)|y \rangle$, and the
corresponding energy dependent Green's function at the negative (i.e. below the
threshold) energy $E=-\kappa^2/\mu$:
\beq
G\left(x,y; -{\kappa^2 \over \mu} \right ) = \int_0^\infty {\cal K}(x, y; \tau)
\, \exp \left( -{\kappa^2 \over \mu} \, \tau \right ) \, d\tau
\label{nrgf}
\eeq
at $x=0$ and $y=0$. We remind that if only the kinetic term is retained in the
Hamiltonian (\ref{hamnr}), i.e. at $\epsilon =0$ and $g=0$, these functions are
${\cal K}_0(0, 0; \tau) = (4 \pi \, \tau /\mu)^{-1/2}$ and $G_0(0,0;
-\kappa^2/\mu)=\mu/(2 \kappa)$. At $\epsilon = 0$ and a nonzero $g$ the Green's
function for the Hamiltonian (\ref{hamnr}) is found as
\beq
G_{\epsilon=0} (0,0; -\kappa^2/\mu) = { G_0(0,0; -\kappa^2/\mu) \over 1- 2 g \,
G_0(0,0; -\kappa^2/\mu)} = { \kappa \over 2 \, \mu} \, {1 \over 1- g \, \mu
/\kappa}~.
\label{gg}
\eeq
The latter expression contains explicitly the pole at $\kappa = \mu \, g$
corresponding to the bound state.

When both the $\epsilon$ and $g$ are nonzero the equation (\ref{gg}) is modified
by replacing the Green's function $G_{\epsilon=0}$ by that for a nonvanishing
$\epsilon$: $G_\epsilon$. The latter Green's function can be expressed in terms
of the corresponding propagator ${\cal K}_\epsilon(0, 0; \tau)$ which can be
found in the textbook \cite{fh}:
\beq
{\cal K}_\epsilon(x, y; \tau) =\sqrt{\mu \over 4 \, \pi \, \tau} \, \exp \left [
-{ \mu \, (x-y)^2 \over 4 \, \tau} + {\epsilon \, (x+y) \over 2} \, \tau +
{\epsilon^2 \over 12 \, \mu} \, \tau^3 \right ]~,
\label{ke}
\eeq
\beq
G_\epsilon \left (0,0; -{\kappa^2 \over \mu} \right ) = \int_0^\infty \sqrt{ \mu
\over 4 \, \pi \, \tau} \, \exp \left( {\epsilon^2 \over 12 \, \mu} \, \tau^3
-{\kappa^2 \over \mu} \, \tau \right ) \, d\tau~,
\label{ge}
\eeq
and the pole position is thus determined from the equation
\beq
2 g \, G_\epsilon \left (0,0; -{\kappa^2 \over \mu} \right ) =1~.
\label{pole}
\eeq
\begin{figure}[ht]
\begin{center}
 \leavevmode
    \epsfxsize=6cm
    \epsfbox{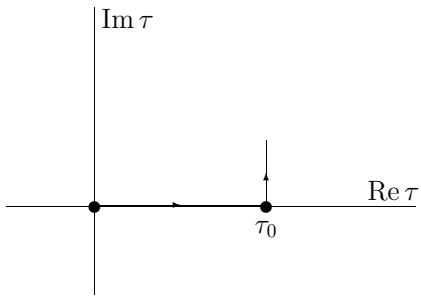}   
    \caption{The contour of integration for the integral in eq.(\ref{ge}). }
\end{center}
\end{figure}

The peculiarity of the latter equation  is that the integral in eq.(\ref{ge}) is
formally divergent, which is the usual situation in a calculation of the energy
of an unstable state. In order to make physical sense, both that energy and the
integral in eq.(\ref{ge}) should be understood as a result of an analytical
continuation in the parameters of the model from the region where the considered
state is stable. In terms of eq.(\ref{ge}) this corresponds to a continuation
from the region of (formally) negative $\epsilon^2$ where the integral is
convergent. The result of such analytical continuation to physical positive
$\epsilon^2$ can be formulated as follows\cite{cc1}: The integration runs along
the real axis of $\tau$ from $\tau =0$ to the value of $\tau$ where the
integrand has minimum, i.e. to $\tau_0=2 \kappa/\epsilon$. From that point the
contour of integration should be turned parallel to the imaginary axis of
$\tau$, corresponding to the direction of the steepest descent (see Fig.2). This
contour rotation gives rise to an imaginary part of the integral, and hence to
an imaginary part of the energy of the resonant state, corresponding to the
decay width of the resonance. Following this procedure one can readily find the
real and the imaginary parts of the integral in eq.(\ref{ge}) and reduce the
equation (\ref{pole}) for the position of the pole to the form
\beq
g \, {\mu \over \kappa} \, \left [ 1 + {i \over 2} \, \exp \left ( - {4 \over 3}
{\kappa^3 \over \mu \, \epsilon} \right ) \right ]   = 1~,
\eeq
which corresponds to the decay rate of the bound state
\beq
\Gamma= 2 \, \mu \, g^2 \, \exp \left ( - {4 \over 3} \, g^3 \,{\mu^2 \over
\epsilon} \right )~.
\label{gf}
\eeq
It can be noted that the exponential factor in this formula is the standard WKB
tunneling exponent in a linear potential, while the pre-exponential factor is a
new result. The described derivation of the formula (\ref{gf}) assumes that the
integral in eq.(\ref{ge}) can be evaluated in the saddle point approximation,
which implies that the parameter in the exponent in eq.(\ref{gf}) is large, i.e.
that $g^3 \, \mu^2 \gg \epsilon$.

\section{Discussion}

In this paper we have refined the calculations of the decay rate
of the boson in the false vacuum and have found the decay rate
of fermion in the false vacuum in (1+1) dimension with the
pre-exponential factor. All calculations, except for the one in Sec.6, have been
performed in
the approximation when the back reaction of the external particle on the
Euclidean bounce solution can be neglected.

The account of the back reaction amount to several new effects which
are different for d=2 and $d>2$. In the (1+1) dimensional case the back reaction
deforms the classical solution which deformation has been described classically
in Ref.\cite{sv} however the calculation of the pre-exponential factor
is beyond our approximation. Such calculation could be
potentially interesting from the stringy perspective. Indeed,
the worldsheet theory on nonabelian string in several models
can be identified with the $CP^N$ model (see \cite{shifman} for a
recent review) which has one true vacuum and a set of metastable
ones. At large N this theory can be treated perturbatively and the
issue of the decay of metastable vacua or in other terns exited strings
can be discussed, Similarly one can discuss the fate the different
excitations on the exited metastable string which is just the problem we have
considered.  In some situations the pre-exponential factor is of the
prime importance since in some range of parameters the
N dependence disappear from the exponent \cite{shifman}. However
it is unclear if the regime with the negligible back reaction could
exist in the worldsheet theory. It seems that the analysis
similar to one in Section 6 could be applicable in this case.

In higher dimensions the situation is more complicated. The point
is that the initial particle evolves along a trajectory in the
complexified Minkowski space\cite{mv3}. At the first stage of the process the
initial particle
"produces" the oscillating bubble in the Minkowski space which
later develops the path in the Euclidean space.   The overlap of the
initial particle and the bubble happens out of the real axis. Let us
also note that there is a possibility of the resonant decay  of the
particle in the false vacuum when the particle mass coincides
with the energy levels of the quantized bubbles in the Minkowski space.

\section*{Acknowledgments}

The work
of A.G. was supported in part by grants CRDF RUP2-261-MO-04
and RFBR-04-011-00646 .
A.G. thanks FITP Institute at University of Minnesota
where the part of the work   has been done   for the kind hospitality and
support.  He also thanks KITP at UCSB
for the hospitality during the program "Mathematical Structures
in  String Theory" supported by grant NSF PHY99-07949.
The work of MBV is supported in part by the DOE grant
DE-FG02-94ER40823.

\end{document}